\documentclass[preprint,showpacs,preprintnumbers,chaos]{revtex4}
\usepackage[dvips]{epsfig}

\newcommand{\be}{\begin{equation}}
\newcommand{\ee}{\end{equation}}
\newcommand{\bea}{\begin{eqnarray}}
\newcommand{\eea}{\end{eqnarray}}

\begin{document}

\title{Chaotic and Pseudochaotic Attractors of Perturbed Fractional 
Oscillator}
\author{\ G.M. Zaslavsky\\\textit{{\small Courant Institute of
Mathematical Sciences, New York University}}
\\\textit{{\small 251 Mercer St., New York, NY 10012}}\\\smallskip{}
\textit{{\small and Department of Physics, New York 
University}}\\\textit{{\small 2-4 Washington Place, New York, NY 10003}}
\\\ \\ A.A. Stanislavsky\\\textit{{\small Courant Institute 
of Mathematical Sciences, New York University}}
\\\textit{{\small 251 Mercer St., New York, NY 10012}}\\\smallskip{}
\textit{{\small and Institute~of~Radio~Astronomy,
Ukrainian~National~Academy~
of~Sciences}}\\\textit{{\small4~Chervonopraporna~St., Kharkov~61002,
Ukraine}}
\\\ \\ M. Edelman \\textit{{\small Courant Institute of
Mathematical Sciences, New York University}}
\\\textit{{\small 251 Mercer St., New York, NY 10012}}
}

%\author[label1,label2]{George~M.~Zaslavsky\corauthref{cor1}}\corauth[cor1]{Corresponding  author. E-mail: zaslav@cims.nyu.edu}
%\author{, \ \ }
%\author[label1,label3]{Aleksander~A.~Stanislavsky\thanksref{lab1}}
%\author{, \ \ }
%\thanks[lab1]{E-mail: alexstan@ira.kharkov.ua}
%\author[label1]{Mark~Edelman\thanksref{lab2}}
%\author{, \ \ }
%\thanks[lab2]{E-mail: edelman@cims.nyu.edu}

%\address[label1]{Courant~Institute~of~Mathematical~Sciences, New~York~University, 
%251~Mercer~St., New~York, NY~10012, USA}

%\address[label2]{Department~of~Physics, New~York~University, 2-4~Washington~Place,\\ 
%New~York, NY~10003, USA}

%\address[label3]{Institute~of~Radio~Astronomy, Ukrainian~National~Academy~of~Sciences, 
%4~Chervonopraporna~St., Kharkov~61002, Ukraine}

%Date: July 22

\begin{abstract}
We consider a nonlinear oscillator with fractional derivative of the order 
$\alpha$. 
Perturbed by a periodic force, the system exhibits chaotic
motion called fractional chaotic attractor (FCA). The FCA is compared to
the ``regular'' chaotic attractor. The properties of the FCA are discussed
and the ``pseudochaotic'' case is demonstrated.
\end{abstract}

%\begin{keyword}

%Chaotic attractor \sep
%Fractional oscillator \sep Pseudochaos

%\end{keyword}

%\end{frontmatter}


\maketitle

\section{Introduction}
\thispagestyle{empty}
It became evident now that the random dynamics can appear
as an intrinsic property of real systems and that different kind of
randomness represents a level of complexity of motion.
While the Hamiltonian chaos is relevant to the Hamiltonian systems, chaotic
properties of the dissipative dynamics are very different revealing the
chaotic attractors, strange non-chaotic attractors, quasi-attractors,
etc. \cite{Ott,Zaslavsky1978,Alligood}. 
In this paper we would like to describe one more way
of the occurrence of chaotic or pseudo-chaotic attractors in dissipative
systems. Namely, the forced system is a fractional nonlinear oscillator
(FNO), i.e. a nonlinear oscillator with a fractional derivative of the
order  $0< \alpha <2$ with respect to time instead of the second order
derivative. 

Although the fractional calculus has more than few-hundred-year story,
its application to the contemporary physics is very recent and, mainly,
it is related to the complexity of the media in classical and quantum
treatment. Let us mention only a few of them: fractional kinetics
\cite{37,rz1,Piryatinska}, 
wave propagation in a media with fractal properties 
\cite{rz2,rz8,rz9}, nonlinear optics \cite{Weitzner}, quantum mechanics 
\cite{rz3}, quantum field theory \cite{rz4}, and many others.The formal
issues related to the contemporary fractional calculus are well reflected
in the monographs  \cite{Gelfand,rz5,rz6,PodlubnyBook,47}. There are different
possibilities for interpretation of fractional derivatives. Let us mention
the probabilistic interpretation  \cite{rz6,Machado,rz7,Stanislavsky2004}
and the relation to a
dissipation  of the considered system 
\cite{41,Gorenflo,34,StanislavskyPRE04}. 
The latter one will be related to our
paper. 

The paper contains some necessary definitions in Sec.~\ref{par3}
and in two appendices. In Sec.~\ref{par5} we demonstrate how the
fractional derivative can be related to a dissipation in the system. The 
Sec.~\ref{par6} is devoted to the main object of the paper:
fractional chaotic attractor (FCA) and some of its features. We speculate
on the existence of fractional ``pseudochaotic'' attractor,
i.e. dissipative random dynamics with zero Lyapunov exponent.

\section{Definitions}\label{par3}
In this section we put some necessary definitions. The left
$_aD^{\alpha}_t$ and right $_tD^{\alpha}_b$ Riemann-Liouville fractional
derivatives of order $\alpha$ are defined as
$$
_aD^{\alpha}_tf(t)={\frac{1}{\Gamma(n - \alpha)}} {\frac{d^n}{dt^n}} \int^t_a(t-
      \tau)^{n- \alpha -1} f(\tau ) d \tau\ , \
$$
\begin{equation}
%\ \ \ \ \ \ \ \ 
%\ \ \ \ \ \ \ \
_tD^{\alpha}_bf(t)={\frac{(-1)^n}{\Gamma(n - \alpha)}} {\frac{d^n}{dt^n}}
\int^b_t(\tau - t)^{n- \alpha -1} f(\tau ) d \tau\ , \label{eqN1}
\end{equation}
$$
(n-1 \le \alpha \le n)  \ . \
$$
To construct a solution for a process described by an equation with
fractional derivatives, one needs the initial conditions that can be
written as  
\begin{equation}
\lim_{t\rightarrow a} {_aD^{\alpha -k}_t} f(t) = a_k  \ , \ \ \ \ \ (k=1,2,...,n)
\label{eqN2}
\end{equation}
or
\begin{equation}
\lim_{t\rightarrow b} {_tD^{\alpha -k}_b} f(t) = b_k  \ , \ \ \ \ \
(k=1,2,...,n)  \ . \
\label{eqN3}
\end{equation}
These conditions may have no physical meaning (for a detailed discussion
see \cite{PodlubnyBook,rz10}).
The important feature of the derivatives  (\ref{eqN1})
is that they have no symmetry with respect to the time reflection 
$t \rightarrow -t$.

In the following we use the so-called Caputo derivative 
\cite{Rabotnov,Caputo}
defined as
$$
_a^CD^{\alpha}_tf(t)={\frac{1}{\Gamma(\alpha - n)}}  \int^t_a(t-
      \tau)^{n- \alpha -1} f^{(n)}(\tau ) d \tau\ , \
$$
\begin{equation}
%\ \ \ \ \ \ \ \ \ \ \ \ \ \ \ \
_t^CD^{\alpha}_bf(t)={\frac{(-1)^n}{\Gamma(\alpha -n)}} 
\int^b_t(\tau - t)^{n- \alpha -1} f^{(n)}(\tau ) d \tau\ , \label{eqN4}
\end{equation}
$$
(n-1 \le \alpha \le n)  
$$
with the regular type of initial conditions
\begin{equation}
f^{(k)}(a) = a_k  \ \ {\rm or} \ \ f^{(k)}(b) = b_k  \ ; \ \ \ \ \
k=0,1,...,(n-1) \ .\
\label{eqN5}
\end{equation}
For the left Caputo derivative we use a notation
\begin{equation}
_0^CD^{\alpha}_t \equiv D^{\alpha } \ .\
\label{eqN6}
\end{equation}

Left fractional oscillator will be described by the equation:
\begin{equation}
_0D^{\alpha}_t x_1(t)-\lambda x_1(t)=0\,,\label{eq31}
\end{equation}
with initial conditions (\ref{eqN2}).
Solution of (\ref{eq31})  can be found with the help of the Laplace transform
\cite{PodlubnyBook}:
\begin{equation}
s^\alpha X_1(s)-\lambda X_1(s)=\sum_{k=1}^n a_ks^{k-1}\,,\label{eq32}
\end{equation}
where 
\begin{equation}
X_1(s)=\int^{\infty}_0 e^{-st} x_1(t) dt   \ . \label{eqN7}
\end{equation}
It follows from (\ref{eq32}) 
\begin{equation}
X_1(s)=\sum_{k=1}^n a_k\frac{s^{k-1}}{s^\alpha-\lambda}\  \label{33}
\end{equation}
and the inverse Laplace transform gives 
\begin{equation}
x_1(t)=\sum_{k=1}^n a_k\,t^{\alpha-k}\,E_{\alpha,\,\alpha-k+1}
(\lambda t^\alpha)\ , \ \ \ \ (t \ge 0) \ , \    \label{eq34}
\end{equation}
where 
\begin{equation}
E_{\alpha,\beta}(z)=\sum_{k=0}^\infty \frac{z^k}{\Gamma(\alpha k+\beta)}\,,\label{eq35}
\end{equation}
is the two-parameter Mittag-Leffler function \cite{Erdelyi}. This equation describes a causal
evolution of the system from the present to the future.

The right fractional oscillator is given by the equation
\begin{equation}
_tD^{\alpha}_0 x_2(t)-\lambda x_2(t)=0\,,\label{eq36}
\end{equation}
with the initial conditions (\ref{eqN3}).
The Laplace transform
in this case 
\begin{equation}
X_2(p)=\int^0_{-\infty} e^{pt}\,x_2(t)\,dt\ , \ \ \ (t \le 0)  \label{eq37}
\end{equation}
gives
\begin{equation}
p^\alpha X_2(p)-\lambda X_2(p)=\sum_{k=1}^n a_kp^{k-1}\,.\label{eq39}
\end{equation}
with the corresponding anti-causal solution
\begin{equation}
x_2(t)=\sum_{k=1}^n a_k\,(-t)^{\alpha-k}\,E_{\alpha,\,\alpha-k+1}
[\lambda (-t)^\alpha]\ ,  \ \ \ (t \le 0) \ . \   \label{eq311}
\end{equation}
Similar results can be given for the equations with left and  
right Caputo derivatives.

\section{Decay rate analysis}\label{par5}

In this section we analyze the dynamics of a nonlinear fractional
oscillator  (see also \cite{Chatterjee}).

\begin{figure}
\centering
\includegraphics[width=12 cm]{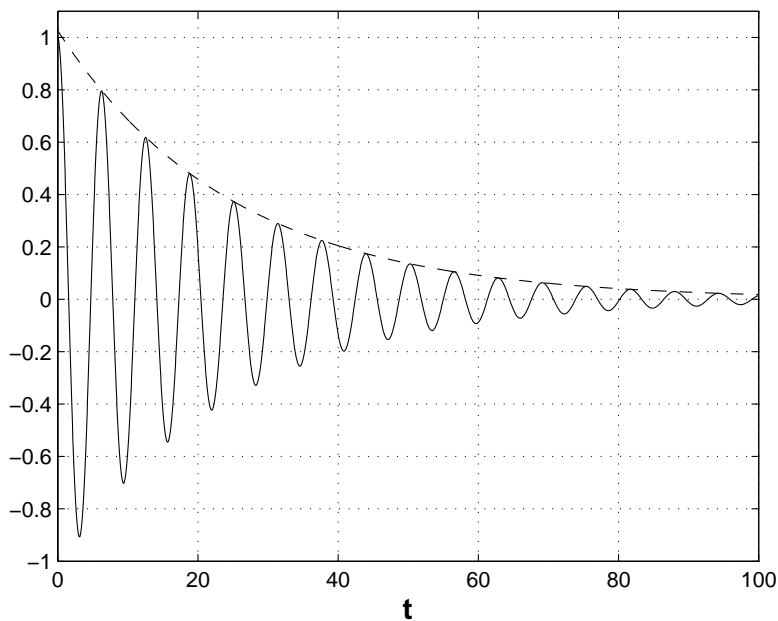}
\caption{\label{fig1a}Rate decay of the linear fractional oscillation 
with initial conditions $x(0)=1$, $x'(0)=0$ and the index $\alpha=1.95$,
$f_{1.95}(0)=-0.0256$. Solid line corresponds to the solution (\ref{eq52}),
dash-line - to the approximation  (\ref{eq58})}
\end{figure}

Let us start from the linear fractional oscillation satisfying
the equation
\begin{equation}
D^\alpha x+x=0\,, \qquad\qquad (1<\alpha<2) \ , \  \label{eq51}
\end{equation}
where $ D^\alpha$ is  the Caputo left fractional 
derivative  (see  (\ref{eqN6})).
Equation (\ref{eq51}) has a solution in a form of  one-parameter Mittag-Leffler function 
\begin{equation}
x(t)=AE_\alpha(-t^\alpha)=A\sum_{k=0}^\infty\frac{(-1)^kt^{\alpha k}}{\Gamma(\alpha k+1)}=AE_{\alpha,1}(-t^\alpha)\label{eq52}
\end{equation}
provided that $x'(0)=0$ and $x(0)=A$ (see Appendix 1 for details). 
According to \cite{Gorenflo,StanislavskyPHYSA05}, the Mittag-Leffler function may be decomposed into two terms
\begin{equation}
E_\alpha(-t^\alpha)=f_\alpha(t)+g_\alpha(t)\,,\label{eq53} 
\end{equation}
where
\begin{equation}
\ \ \ \ \ \ \ \ \ \ \ \ \ \ \ \
f_\alpha(t) = \frac{1}{\pi}\int_0^\infty e^{-\,rt}\,\frac{r^{\alpha-1}
\sin(\pi\alpha)}{r^{2\alpha}+2r^\alpha\cos(\pi\alpha)+1}\,dr\,,
\label{eq54}\\ 
\end{equation}
$$
g_\alpha(t) = \frac{2}{\alpha}\,e^{\,t\cos(\pi/\alpha)}\cos
\left[t\sin\Bigl(\frac{\pi}{\alpha}\Bigr)\right].
$$
The first term is determined by a cut on the complex plane for  
Mittag-Leffler function, and the 
second one is related to the poles. For $\alpha=2-\varepsilon$ ($\varepsilon\ll 1$) 
we can use an expansion over $\varepsilon$:
\begin{eqnarray}
f_\alpha(t)&\approx&-\varepsilon\int_0^\infty
e^{-rt}\frac{r\,dr}{(r^2+1)^2}\nonumber\\ & & = -\frac{\varepsilon}{2}
\Bigl\{1-t[\mathrm{ci}(t)\sin t-\mathrm{si}(t)\cos t]\Bigr\}\,,\qquad
\varepsilon\ll 1\, \label{eq56} 
\end{eqnarray}
where $\mathrm{ci}(t)$ and $\mathrm{si}(t)$ are  sine and cosine
integral functions respectively.
For large $t$ 
the function $E(-t^\alpha)$ has  algebraic asymptotics 
of $f_\alpha(t)$ \cite{Gorenflo,Erdelyi}
\begin{equation}
f_{2-\varepsilon}(t)\sim\frac{t^{-2+\varepsilon}}{\Gamma(\varepsilon-1)}\,,\qquad\ t\gg t^*=\frac{12}{\pi\varepsilon}\ln\frac{1}{\varepsilon}\,,\qquad\varepsilon\ll 1\,,\label{eq57}
\end{equation}
i.e. the first term in (\ref{eq53})
decays algebraically in time while the 
second one decays exponentially. 
The decay rate of $g_\alpha(t)$ is defined by its amplitude
\begin{equation}
\max g_\alpha(t)=\frac{2}{\alpha}\,e^{-\,\gamma t}\,,\label{eq58}
\end{equation}
where 
\begin{equation}
\gamma=-\cos(\pi/\alpha) \ . \  \label{eq25a} 
\end{equation}
It is useful to remark that
$0<\gamma<1$ for $1<\alpha<2$, and $\gamma\approx\pi\varepsilon/4$
for $\varepsilon\ll 1$. For a comparison  the function
$E(-t^\alpha)$ and its envelope  (\ref{eq58}) are presented in
Fig.~\ref{fig1a}. 

The analysis can be extended to a nonlinear fractional oscillator since
for small $\varepsilon $ one can apply the averaging over fast oscillations.

As an example, consider the fractional Duffing equation
\begin{equation}
D^\alpha y-y+ay^3=0\,,\qquad\qquad 1<\alpha<2,\label{eq59}
\end{equation}
where $a > 0$ is a constant. 
The steady states are: $y=0$, (unstable) and $y=\pm 1/\sqrt{a}$ (stable).

\begin{figure}
\centering
\includegraphics[width=12 cm]{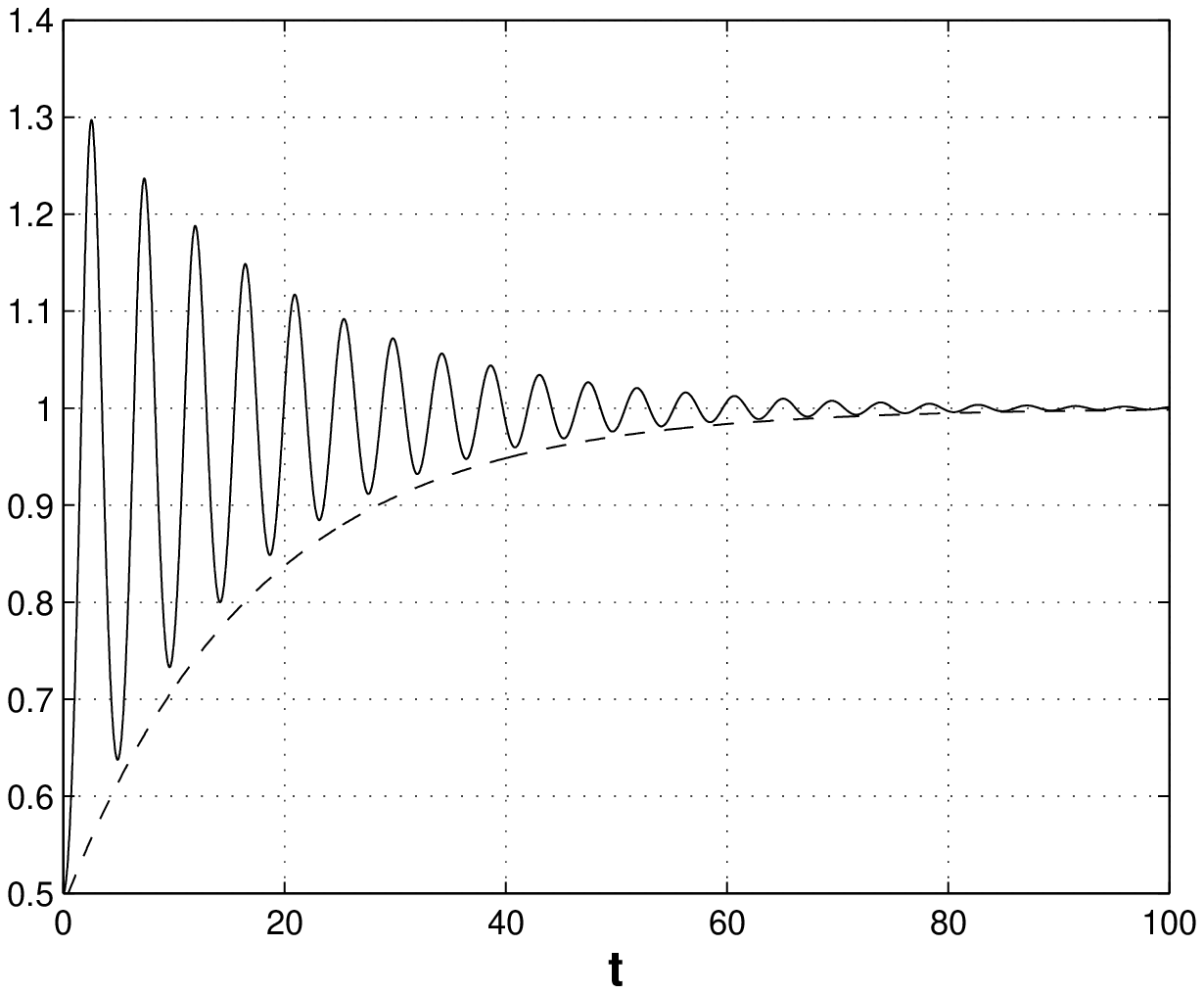}
\caption{\label{fig1}Rate decay of nonlinear oscillations described by 
the fractional Duffing equation with initial conditions $x(0)=0.5$, $x'(0)=0$ 
and the index $\alpha=1.95$. The solid line is a 
numerical solution of (\ref{eq59}), the dashed line is the envelope 
curve according to (\ref{eq512a}).}
\end{figure}

Consider Eq. (\ref{eq59}) near a stable fixed point
by the change $y\to w+1/\sqrt{a}$. 
Then 
\begin{equation}
D^\alpha w+2w+3\sqrt{a}w^2+aw^3=0\,.\label{eq511}
\end{equation}
Close to the stable fixed point we have a linear equation
\begin{equation} 
D^\alpha w_0+2w_0=0\label{eq511a}
\end{equation}
with a solution
\begin{equation}
w_0(t)=BE_\alpha(-2t^\alpha)\,,\label{eq512}
\end{equation}
and $B$ is a constant.
For $\alpha=2-\varepsilon$ and $\varepsilon \ll 1$
expression (\ref{eq512}) is well approximated by the relation
\begin{equation}
w_0(t)\approx\frac{2B}{2-\varepsilon}e^{\,-\sqrt{2}\gamma t}\cos(\sqrt{2}\,t)+\mathit{O}(\varepsilon^2).\label{eq512a}
\end{equation}
with $\gamma $ from  (\ref{eq25a}).
Fig.~\ref{fig1} shows a numerical simulation of Eq.(\ref{eq59}) in comparison with the amplitude obtained from (\ref{eq512a}). The numerical analysis of Eq. (\ref{eq59}) is based on the algorithms described in \cite{Diethelm2005}.

When $\alpha=2$, Eqs. (\ref{eq59}) and (\ref{eq511}) become undamped. The
 leading term of the frequency of the oscillation $w$ is
 $\omega_0=2^{1/\alpha}$. 
From \cite{Landau}, a nonlinear correction to this frequency is
\begin{equation}
\omega_1=-\frac{3\,aB^2}{2\sqrt{2}}<0,\label{eq13}
\end{equation}
when $aB^2 \ll 1$, i.e. $|\omega_1| \ll \omega $.

From Eq. (\ref{eq512a}), we can present $w$ in the form
\begin{equation}
w(t)\approx w_1(t) + \delta w(t) = Be^{-\sqrt{2}\gamma
  t}\cos\Bigl[(\omega_0+\omega_1)t+\beta\Bigr] +  \delta w(t)\ ,\label{eq514}
\end{equation}
where $\beta$ is constant and $ \delta w(t) $ is a correction to  
$\omega_1(t)$ due to  the term
$f_{\alpha }(t)$ in (\ref{eq57}) that describes the polynomial decay of
oscillations for fairly large $t$.

Concluding this section one state that the fractional generalization of
the considered nonlinear oscillations is reduced to some effective decay
of the oscillations similarly to the fractional linear oscillator. The
rates of the decay can be estimated and, roughly speaking, the larger is
the deviation $\varepsilon = 2 - \alpha $, the stronger is the decay.

\section{Fractional Chaotic Attractor (FCA)}\label{par6}

It is well known that periodic force applied to a nonlinear oscillator,
for general situation, leads to the Hamiltonian chaotic dynamics in some
part of phase space, while the same problem with dissipation can lead to the
chaotic attractor. One can expect that the periodically perturbed
fractional nonlinear oscillator should display a kind of chaos that we
call FCA. Description of this phenomenon is the subject of this
section. The basic equation is
\begin{equation}
D^\alpha x-x+x^3=F\sin(\nu t)\,,\qquad\qquad 1<\alpha<2,\label{eq516}
\end{equation}
where $F$ and $\nu$ are parameters of the perturbation. For fairly small 
$\varepsilon = 2-\alpha$,
let us introduce  
a cojoint equation

%$$
% \ddot{x}+2\sqrt{2}\gamma\dot{x}-x-c+x^3=F_1\sin(\nu_1 t)\ ; \ \ \ \gamma = \pi
%\varepsilon /4 \ , \ \ \ c = (\sqrt{2}/2) \varepsilon \ ,
%$$
%\begin{equation}
%F_1=F(1+(3\ln 2 /4) \varepsilon)\ ; \ \ \ \nu_1=\nu (1+(\ln 2 /4)  \varepsilon)\ . \ \label{eq517}
%\end{equation}

\begin{equation}
\ddot{x} + a_1 \dot{x}-x+x^3=F_1\sin(\nu_1 t)\ \ . \ \label{eq517}
\end{equation}

\begin{figure}
\centering
\includegraphics[width=14 cm]{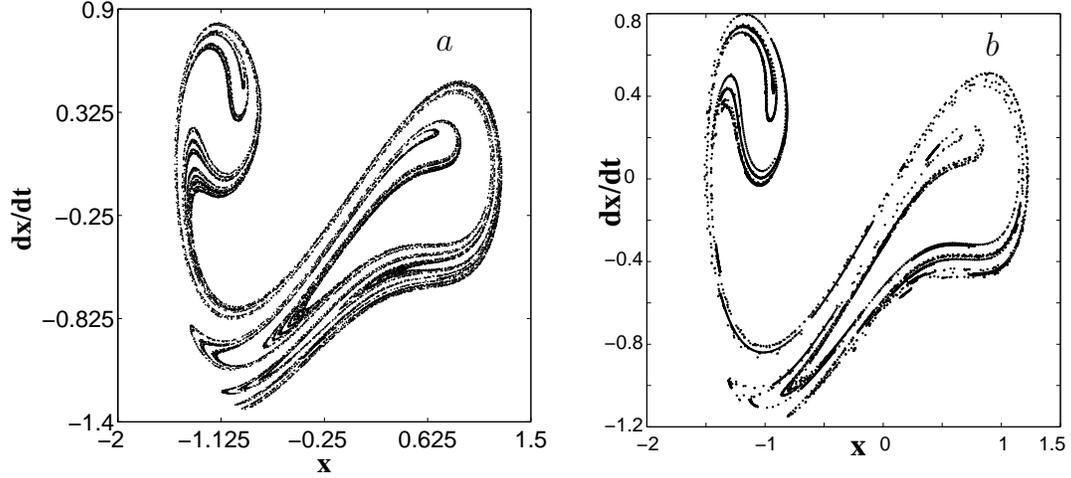}
\caption{\label{fig6} Poincar$\acute{\rm e}$ maps: (a) for the ordinary
  Duffing equation driven by periodic force: $a_1=0.1172$, $F_1=0.279$, 
$\nu_1=1.0149$; 
(b) for the fractional Duffing
 equation 
with the index $\alpha=1.9$, $\nu=1$, $F=0.3$, $t_{max}=16\pi$, driven by periodic force.
Number of trajectories in (b) is  2000.}
\end{figure}

\begin{figure}
\centering
\includegraphics[width=12 cm]{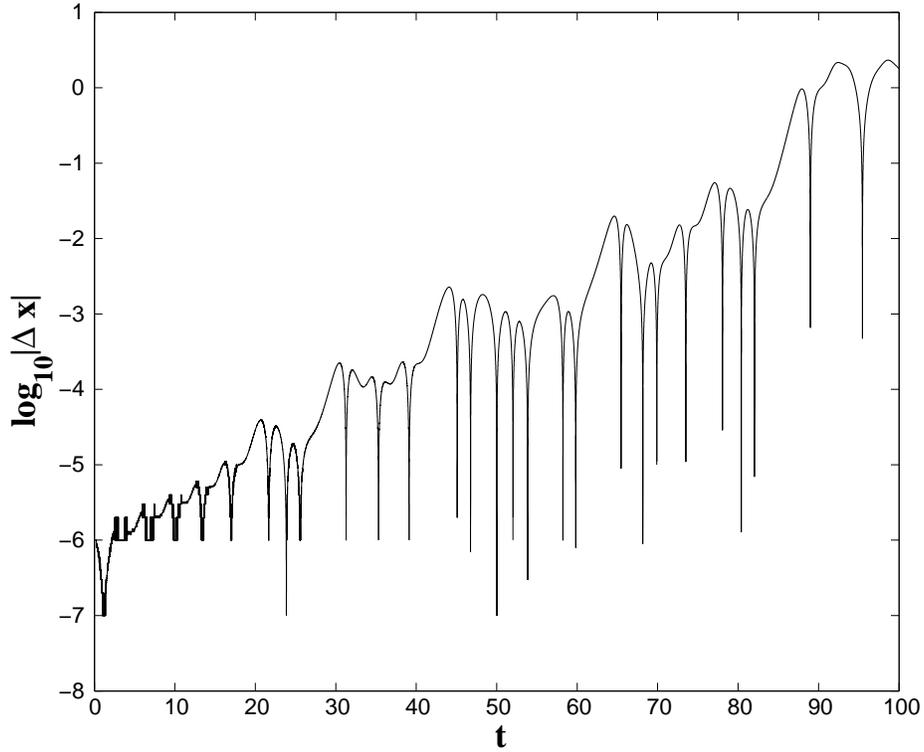}
\caption{\label{figC1} Exponential growth of distance $|\Delta x|$ of 
initially close
  trajectories ($\Delta x= 10^{-6}$, $t=0$) for the FCA. $F=0.3$; $\nu =1$. }
\end{figure}

Simulation of Eq.~(\ref{eq517}) shows a typical chaotic attractor in 
Fig.~\ref{fig6},(a). One can recalculate the value of $\alpha $   into a
corresponding   $a_1$.  As a result, in Fig.~\ref{fig6},(b) we see the
map of the FCA. These figures display a structural difference between CA
and FCA. We can assume that for small $\varepsilon$ and up to the terms of 
$\varepsilon^2$ the behavior of the FCA is similar to the CA of the
cojoint equation. Nevertheless, it seems that this similarity is not
complete and some strong difference is indicated below.

The computational time for  Eq.~(\ref{eq516}) is very large in
order to keep a reasonable accuracy. To reduce this time  one can consider
a short time of computation  but with large number of trajectories . That
is a way how the  Fig.~\ref{fig6},(b) was obtained. It means that
the Poincar\'{e} maps for CA and FCA were obtained in different ways. To
check the presence of chaotic dynamics, we consider  growth of the
distance $| \Delta x | $ for two initially close trajectories for 
Eq.~(\ref{eq516}). This result is in  Fig.~\ref{figC1} and it conforms the
presence of a positive Lyapunov exponent. Nevertheless, we can see almost
regular returns of  $| \Delta x | $ to the initial value $10^{-6}$.

\begin{figure}
\centering
\includegraphics[width=14 cm]{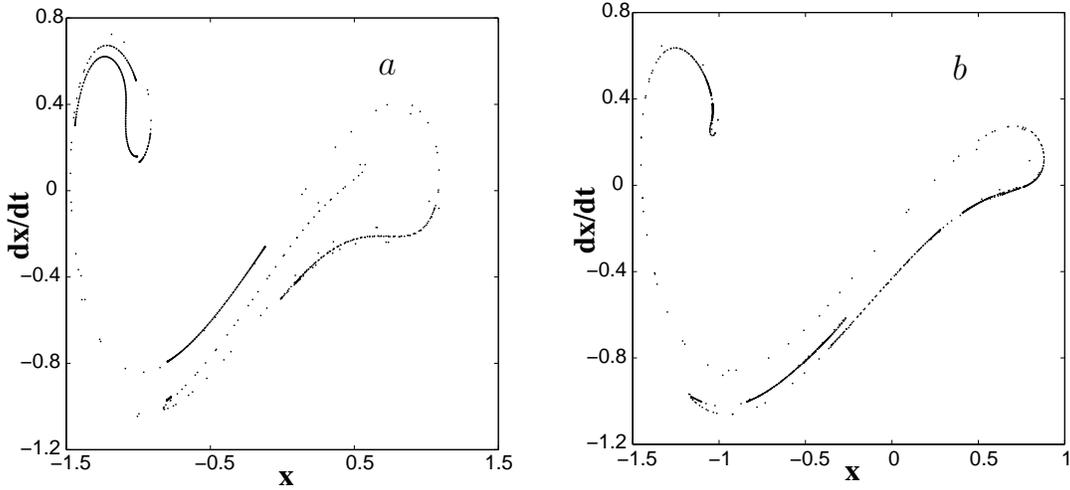}
\caption{\label{figC2} Reduction of structures for the FCA with the growth
  of $\varepsilon = 2 - \alpha $: (a) $\alpha = 1.82$, $\nu=0.985$,
  $F=0.39$; 
(b)  $\alpha = 1.7$, $\nu=0.955$,  $F=0.6$. Number of trajectories is 200.}
\end{figure}

As it was mentioned above, increase of the ``fractionality'' of the the
derivative order $\varepsilon = 2 - \alpha $ can increase the 
dissipation. This
leads to some ``reduction of structures'' in the FCA (see  Fig.~\ref{figC2})
similarly to the case of CA \cite{Zaslavsky1978}.

\begin{figure}
\centering
\includegraphics[width=14 cm]{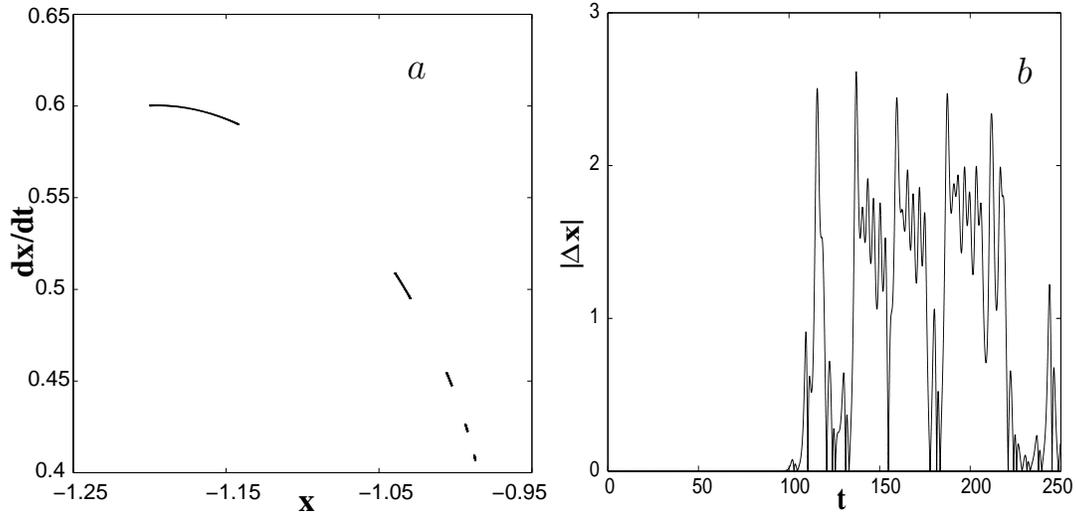}
\caption{\label{figC3} Dying FCA:  $\alpha = 1.82$, $F=0.3$, $\nu=1$;
(a) phase plane of the FCA; (b)  dispersion of initially close two 
trajectories.}
\end{figure}

Finally, we present the case of a ``dying attractor'' \cite{Zaslavsky1978}
in  Fig.~\ref{figC3}. Although the phase portrait looks very regularly,
the dispersion of initially close trajectory shows randomness with the
most probably zero Lyapunov exponent. This case can be related to the
fractional pseudochaotic attractor (FPCA) that has a counterpart
(pseudochaos) in Hamiltonian dynamics  \cite{rz11}.
Two  comments support this hypothesis: (a) there is no separation of the
initially close trajectories for fairly large time after which the
distance jumped to the order one (see  Fig.~\ref{figC3},(b));
(b) there is no structure of the attractor in Fig.~\ref{figC3},(a) even
when we strongly increase a resolution of the  Poincar\'{e} section plot,
i.e. the dimension of the set in  Fig.~\ref{figC3},(a) is rather one than
larger of one as in  Fig.~\ref{fig6},(a).

%\begin{figure}
%\centering
%\includegraphics[width=12 cm]{fig2chaos.eps}
%\caption{\label{fig2}Poincar$\acute{\rm e}$ map for the fractional Duffing equation forced harmonic vibrations
%with the index $\alpha=1.9$ and $t_{max}=16\pi$\,.}
%\end{figure}

%\begin{figure}
%\centering
%\includegraphics[width=12 cm]{fig4chaos.eps}
%\caption{\label{fig4}Poincar$\acute{\rm e}$ map for the ordinary Duffing equation 
%forced harmonic vibrations.}
%\end{figure}

\section{Conclusions}

\begin{figure}
\centering
\includegraphics[width=12 cm]{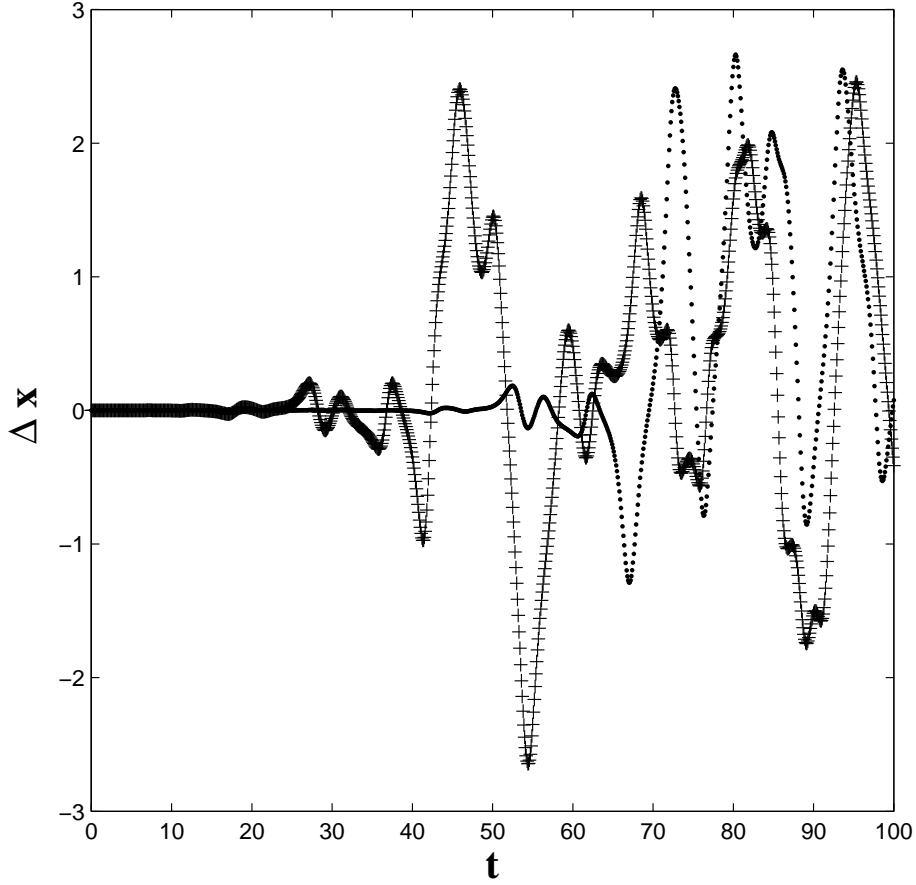}
\caption{\label{fig5}Errors between the numerical calculations with time steps 0.01 and 0.001 (dots);
with time steps 0.001 and 0.1 (pluses)\,.}
\end{figure}

Since the fractional derivatives are time-directed, the equations with
fractional derivatives slightly different from the integer ones by 
$\varepsilon = 2 - \alpha \ll 1 $ can be fairly easy interpreted through
the regular equations with a dissipation. As a result, fractional
nonlinear oscillator behaves like the stochastic attractor in phase space,
being periodically perturbed. The role of the polynomial dissipation is
still elusive. It seems that this term leads to a degradation of the FCA
structure. The error has fast increase with time creating difficulties in
the simulations (see Fig.~\ref{fig5}). 
Due to that we can not provide explicit features of the difference
between CA and FCA.

The resonant case for the linear oscillator can be interpreted in the way
similar to the integer derivative case with $ \alpha = 2$ \cite{Landau}
(see Appendix 2). Our simulations show that applying the fractional
calculus, one can gain a compact formulations of dynamics with new
properties governed by a complexity of the media.

\section*{Acknowledgements}

This work was
supported by the Office of Naval Research, Grants No. N00014-02-1-0056,
U.S. Department of Energy Grant No. DE-FG02-92ER54184, and the NSF Grant 
No. DMS-0417800. Computations were performed at NERSC.
A.S. thanks the Courant Institute of Mathematical Sciences, New York, 
USA, for support and hospitality during the preparation of this work.
A.S. also acknowledges D. Dreisigmeyer for useful discussions.

\section*{Appendix 1}

Here we consider a presentation of the solution to
equation (\ref{eq51}). Two independent solutions are
\begin{equation}
e_\alpha(t)=E_{\alpha,\,1}(-t^\alpha)\,,\qquad i_\alpha(t)=t^{\alpha/2}
E_{\alpha,\,1+\alpha/2}(-t^\alpha)\,.\label{A1}
\end{equation}
Let $u(t)$ be  a function
and $L.T.u(t)$  be its Laplace transform, namely
\begin{equation}
L.T.u(t) = u(s)=\int_0^\infty e^{-st}\,u(t)\,dt.\label{A2}
\end{equation}
Then the Laplace transforms of $e_\alpha(t)$ and $i_\alpha(t)$ become
\begin{equation}
L.T.e_\alpha(t)=\frac{s^{\alpha-1}}{s^\alpha+1}\ ;\ \ \ \ \qquad L.T.i_\alpha(t)=\frac{s^{-1+\alpha/2}}{s^\alpha+1}
\,.\label{A3}
\end{equation}

Let us deform the Bromwich path of integration into the equivalent Hankel
path.
Then the loop starts from minus infinity along the lower side of
negative real axis, encircles $|s|=1$ counter-clockwise and ends at
minus infinity along the upper side of the negative real axis. This allows one to decompose
the functions $e_\alpha(t)$ and $i_\alpha(t)$ into two terms. 
The first contribution arises 
from two borders of the cut along the negative real semi-axis. The second contribution is determined by residues
in the poles $s_0=\exp(j\pi/\alpha)$ and $s_1=\exp(-j\pi/\alpha)$. 
We arrive to the expressions 
(\ref{eq53})-(\ref{eq54}) \cite{Gorenflo} for the function $e_\alpha(t)$. 
Similarly, we can present
\begin{equation}
i_\alpha(t)=h_\alpha(t)+q_\alpha(t)\,,\label{eqA4} 
\end{equation}
where
\begin{equation}
h_\alpha(t)=\frac{1}{\pi}\int_0^\infty e^{-\,rt}\,\frac{r^{-1+\alpha/2}(1-r^\alpha)
\sin(\pi\alpha/2)\,dr}{r^{2\alpha}+2r^\alpha\cos(\pi\alpha)+1}\,,
\label{eqA5} 
\end{equation}
$$
q_\alpha(t)=\frac{2}{\alpha}\,e^{\,t\cos(\pi/\alpha)}\sin
\left[t\sin\Bigl(\frac{\pi}{\alpha}\Bigr)\right].\label{eqA6}
$$
For the detailed  analysis of these expressions see
\cite{StanislavskyPHYSA05}.

\section*{Appendix 2  Driven linear fractional oscillator}

In this section we  discuss briefly some features of linear fractional 
oscillator perturbed by an analog of the resonant
external force (see also \cite{Achar}). ``Free'' and ``forced'' 
oscillations of the fractional oscillator 
depend on the index $\alpha$. The main conclusion is that the dynamical 
response of the driven fractional oscillator is bounded in amplitude for
any relation between the oscillator frequency and the frequency of the
perturbation. 
The finiteness of the response indicates a damped character of fractional 
derivative. In \cite{StanislavskyPRE04} the linear fractional oscillator 
is interpreted as an ensemble average over harmonic oscillations because 
of the interaction of the fractional system with the random environment. 
The intrinsic absorption of the fractional
 oscillator results from the response of each harmonic
oscillator being compensated by an antiphase response of another harmonic 
oscillator (see details in \cite{StanislavskyPRE04}). This shows the main
difference between the resonant phenomenon in regular systems and 
the resonance in fractional systems (both linear and nonlinear).

Let the external force in the linear fractional oscillator 
equation be described by
Mittag-Leffler function
\begin{equation}
D^\alpha x+\omega^2_0x=CE_\alpha(-\omega^2 t^\alpha)\
 ,\qquad\qquad 1<\alpha<2\,,\label{eq61}
\end{equation}
where $C$ is a constant determined by  initial conditions. Since the equation is linear, its solution can be presented in the form similar to \cite{Landau}
\begin{equation}
x(t)=A_1E_\alpha(-\omega_0^2t^\alpha)+\frac{C_1}{\omega^2_0-\omega^2}\left[E_\alpha(-\omega^2t^\alpha)-
E_\alpha(-\omega^2_0t^\alpha)\right]\,,\label{eq62}
\end{equation}
where $A_1$, $C_1$ are new constants. A direct calculation gives 
\begin{equation}
\lim_{\omega\to\omega_0}x(t)=A_1E_\alpha(-\omega_0^2t^\alpha)+
C_1\frac{t^\alpha}{\alpha}\,E_{\alpha,\,\alpha}(-\omega_0^2t^\alpha)\,.\label{eq63}
\end{equation}
For $t\to 0$ the second term tends to zero. For $1<\alpha<2$, $t\gg 1$ we have 
\begin{equation}
t^\alpha E_{\alpha,\,\alpha}
(-\omega^2_0t^\alpha)\sim\frac{t^{-\alpha}}{\omega^4_0\Gamma(1-\alpha)}\,,\label{eq63a}
\end{equation}
and the same is for the first term of (\ref{eq63})
\begin{equation} 
E_\alpha(-\omega_0^2t^\alpha)\sim\frac{t^{-\alpha}}{\omega^2_0\Gamma(1-\alpha)}\,,\quad t\gg 1\,.\label{eq63b}
\end{equation}
Thus, 
the secular term is absent for $1<\alpha<2$.

The resonance is observed only for $\alpha=2$, when Eq. (\ref{eq63}) is 
transformed into the usual
forced harmonic oscillator. Then,   the secular term appears
\begin{eqnarray}
\lim_{\omega\to\omega_0}x_{\alpha=2}(t)&=&A_1E_2(-\omega_0^2t^2)+C_1t^2E_{2,\,2}(-\omega_0^2t^2)\nonumber\\ 
&=&A_1\cos\omega_0 t+\frac{C_1t}{2\omega_0}\sin\omega_0t\,.\label{eq64}
\end{eqnarray} 

Concluding, the linear fractional oscillator forced by a Mittag-Leffler 
oscillation does not go to any resonance for $\omega\to\omega_0$
for $ 1 < \alpha  < 2$.

%\section*{References}


\begin{thebibliography}{2005}
\bibitem{Ott}
E.~Ott, ``Chaos in dynamical systems'', 
Cambridge University Press, New York, 1993.
\bibitem{Zaslavsky1978}
G.~M.~Zaslavsky,  Phys. Lett. A, {\bf 69}, 145-147 (1978).
%\bibitem{WCSH}
%T.~P.~Weldon,  Am. J. Phys. 58 (1990) 936-941;\\ 
%T.~L.~Carroll, Am. J. Phys. 
%63 (1995) 377-379;\\ J.~C.~Sprott,  Am. J. Phys. 68 (2000) 
%758-763;\\ E.~H.~Hellen,  Am. J. Phys. 72 (2004) 499-502.
\bibitem{Alligood}
K.~T.~Alligood, T.~D.~Sauer, J.~A.~Yorke, ``CHAOS: An introduction to
dynamical 
systems'', Springer-Verlag, New York, 1996.
\bibitem{37}
G.~M.~Zaslavsky,  Physica D, {\bf 76}, 110-122 (1994); Chaos, {\bf 4}, 
25-33 (1994).  
\bibitem{rz1} 
G.~M.~Zaslavsky,  Phys. Reports, {\bf 371}, 461 (2002).
\bibitem{Piryatinska}
A.~Piryatinska, A.~I.~Saichev, W.~A.~Woyczynski, 
Physica A, {\bf 349}, 375-420 (2005).
\bibitem{rz2}
T.~L.~Szabo, J. Acoust. Soc. of Am., {\bf 96}, 491 (1994); 
D.~T.~Blackstock, ibid. {\bf 77}, 2050 (1985); 
W.~Chen, S.~Holm, ibid. {\bf 115}, 1424 (2004).
\bibitem{rz8}
R.~R.~Nigmatullin, Phys. Sta. Sol. (b), {\bf 133}, 425 (1986). 
\bibitem{rz9}
T.~F.~Nonnenmacher, In ``Lecture Notes in Physics'', v. 381, p. 309,
Springer, Berlin, 1991.
\bibitem{Weitzner}
H.~Weitzner, G.~M.~Zaslavsky, Commun. Nonlin. Sci. 
and Numer. Simul., {\bf 8}, 273-281 (2003).
\bibitem{rz3}
N.~Laskin, Phys.Rev. E, {\bf 66}, 056108 (2002).
\bibitem{rz4}
E.~Goldfain, Chaos, Solitons and Fractals,  {\bf 19}, 1023-1030 (2004);
{\bf 22}, 513-520 (2004);
{\bf 23}, 701-710 (2005).
\bibitem{Gelfand}
I.~M.~Gelfand, G.~E.~Shilov, ``Generalized Functions I: 
Properties and Operations'',
Academic, New York, 1964.
\bibitem{rz5}
S.~G.~Samko, A.~A.~Kilbas, O.~I.~Marichev, ``Fractional Integrals and
Derivatives and Some of Their Applications'', Nauka i Technika, Minsk
(1987) (in Russian).
\bibitem{rz6}
M.~M.~Meerschaert, D.~A.~Benson, H.~P.~Scheffler, B.~Baeumer,
Phys.Rev. E, {\bf 66}, 060102 (2002).
\bibitem{PodlubnyBook}
I.~Podlubny, ``Fractional Differential Equations'', Academic Press, 
New York, 1999.
\bibitem{47}
K.~S.~Miller, B.~Ross, ``An Introduction to the Fractional Calculus
and Fractional Differential Equations'', Wiley, New York, 1993.
%\bibitem{Barnett}
%A.~H.~Barnett,  Dissipation in Deforming Chaotic Billiards, 
%PhD thesis, Courant Institute of
%Mathematical Sciences, New York, 2000.
%\bibitem{Grigorenko}
%I.~Grigorenko, E.~Grigorenko,  Phys. 
%Rev. Lett. 91 (2003) 034101.
%\bibitem{Podlubny}
%I.~Podlubny,  
%Fract. Calc. Appl. Anal. 5 (2002) 367-386.
\bibitem{Machado}
A.~I.~Saichev and G.~M.~Zaslavsky, Chaos, {\bf 7}, 753-764 (1997).
%J.~A.~T.~Machado, Fract. 
%Calc. Appl. Anal., {\bf 8},  73-80 (2003). 
%\bibitem{Kempfle}
%S.~Kempfle, I.~Sch$\ddot{\rm a}$fer, H.~Beyer, Nonlinear Dyn. 29 (2002)
%99-127.
%\bibitem{Chechkin}
%A.~V.~Chechkin, V.~Yu.~Gonchar, M.~Szydlowsky, Phys. of Plasmas 9 (2002)
%78-88.
\bibitem{rz7}
V.~V.~Uchaikin, V.~M.~Zolotarev, ``Chance and Stability. Stable
Distributions and their Applications'', VSP, Utrecht (1999).
\bibitem{Stanislavsky2004}
A.~A.~Stanislavsky, Theor. and Math. 
Phys., {\bf 138},  418-431 (2004).
\bibitem{41}
F.~Mainardi, 
Chaos, Soliton \& Fractals, {\bf 7}, 1461-1477 (1996).
\bibitem{Gorenflo}
R.~Gorenflo, F.~Mainardi, In Proceedings of the International 
Workshop on the Recent 
Advances in Applied Mathematics (RAAM '96), Kuwait, 1996, 
Kuwait University, Department of Mathematics and Computer Science, 
pp. 193-208. 
\bibitem{34}
M.~Seredy$\acute{\rm n}$ska, A.~Hanyga, J. Math. Phys., {\bf 41}, 
(2135-2156 (2000).
\bibitem{StanislavskyPRE04}
A.~A.~Stanislavsky,  Phys.Rev. E, {\bf 70}, 051103 (2004).
%\bibitem{Meerschaert}
%M.~M.~Meerschaert, H.-P.~ Scheffler, J. Appl. Probab. 41 (2004) 623-638.
%\bibitem{Duffing}
%G.~Duffing, Erzwungene Schwingung bei ver$\ddot{\rm a}$nderlicher 
%Eigenfrequenz und ihre technische 
%Bedeutung, Vieweg, Braunschweig, 1918.
%\bibitem{Huberman}
%B.~A.~Huberman, J.~P.~Crutchfield,  Phys. Rev. 
%Lett. 43 (1979) 1743-1747.
%\bibitem{Jones}
%B.~K.~Jones, G.~Trefan, Am. J. Phys. 69 (2001) 464-469.
%\bibitem{Mahmoud}
%G.~M.~Mahmoud, A.~A.~Mohamed, S.~A.~Aly,  Physica A 292 (2001) 193-206.
%\bibitem{Dreisigmeyer}
%D.~W.~Dreisigmeyer, P.~M.~Young, J. Phys. A.: Math. Gen. 36 (2003)
%8297-8310;\\
%D.~W.~Dreisigmeyer, P.~M.~Young, 
%J. Phys. A.: Math. Gen. 37 (2004) 
%L117-L121;\\ D.~W.~Dreisigmeyer, P.~M.~Young, 
%arXiv:physics/0402056 
%(2004) 1-14.
\bibitem{rz10}
M.~D.~Ortigueria, Signal Processing, {\bf 83}, 2301 (2003).
%\bibitem{Bateman}
%H.~Bateman,  Phys. Rev., {\bf 38},  815-819 (1931).
%\bibitem{Zakharov}
%V.~E.~Zakharov, E.~A.~Kuznetsov, 
%Physics-Uspekhi, {\bf 40}, 1087-11 (1997).
\bibitem{Rabotnov}
Yu.~Rabotnov, ``Creep problems in structural members'', North-Holland,
Amsterdam, 1969, p.~129. Originally published in Russian as: {\it
Polzuchest' Elementov Konstruktsii}, Nauka, Moscow, 1966.
\bibitem{Caputo}
M.~Caputo,  J. Roy. Astron. Soc., {\bf 13}, 529-539 (1967). 
\bibitem{Erdelyi}
A.~Erd$\acute{\rm e}$lyi, ``Higher Transcendental Finctions'', 
Vol. III, Sec. 18, McGraw-Hill, New York, 1955.
\bibitem{Chatterjee}
A.~Chatterjee,  Nonlinear Dyn., {\bf 32}, 
323-343 (2003);\\ P.~Wahi, A.~Chatterjee,  Nonlinear Dyn., 
{\bf 38}, 3-22 (2004).
\bibitem{StanislavskyPHYSA05}
A.~A.~Stanislavsky,  Physica A, {\bf 354}, 101-110 (2005).
\bibitem{Diethelm2005}
K.~Diethelm, N.~J.~Ford, A.~D.~Freed, Yu.~Luchko, 
Comput. Methods Appl. Mech. Engrg., {\bf 194}, 743-773 (2005).
\bibitem{Landau}
L.~D.~Landau, E.~M.~Lifschitz, ``Mechanics'', 3rd ed., 
Pergamon Press, Oxford, England, 1976.
\bibitem{rz11}
O.~Lyubomudrov, M.~Edelman, G.~M.~Zaslavsky,
Int. J. Mod. Phys. B, {\bf 17}, 4149 (2003);
 G.~M.~Zaslavsky, M.~Edelman, Physica D, {\bf 193}, 128-147 (2004).
\bibitem{Achar}
B.~N.~Narahari Achar, J.~W.~Hanneken, T.~Clarke,  
Physica A, {\bf 309}, 275-288 (2002);
B.~N.~Narahari Achar, J.~W.~Hanneken, T.~Clarke, Physica A, 
{\bf 339}, 311-319, (2004).
%\bibitem{Bauer}
%P.~S.~Bauer,  Proc. Natl. Acad. Sci. 17 (1931) 311-314.
%\bibitem{Riewe}
%F.~Riewe, Phys. Rev. E 53 (1996) 1890-1899;\\ 
%F.~Riewe, Phys. Rev. E 55 (1997) 3581-3592.
%\bibitem{Agrawal}
%O.~P.~Agrawal, J. Math. Anal. 
%Appl. 272 (2002) 368-379.
%\bibitem{Baleanu}
%D.~Baleanu, T.~Avkar, Nuovo Cimento 119 (2004) 
%73-79;\\ D.~Baleanu,  in: Proceedings of 1st IFAC 
%Workshop on Fractional Differentiation and its Applications, 
%Bordeaux, France, 
%July 19-21, 2004, pp.~597-602;\\ S.~I.~Muslih, D.~Baleanu, 
%J. Math. Anal. Appl. 304 (2005) 599-606.
%\bibitem{35}
%M.~Klimek, Czech. J. Phys. 51 (2001) 
%1348-1354;\\ M.~Klimek, Czech. J. Phys. 52 (2002) 1247-1253.
%\bibitem{36}
%T.~T.~Hartley, C.~F.~Lorenzo, H.~K.~Qammar,  IEEE 
%Transactions on Circuits and Systems 42 (1995), 
%485-490;\\ C.~F.~Lorenzo, T.~T.~Hartley, 
%Nonlinear Dyn. 29 (2002) 57-98;\\ 
%T.~T.~Hartley, C.~F.~Lorenzo, 
%Nonlinear Dyn. 29 (2002) 201-233.
%\bibitem{38}
%S.Nimmo, A.~K.~Evans,  Chaos, Solitons \& Fractals 10 (1999) 1111-1118.
%\bibitem{39}
%G.~Turchetti, D.~Usero, L.~Vazquez, Tamsui 
%Oxford J. of Math. Sci. 18 (2002) 31-44.
%\bibitem{40}
%Xin Gao, Juebang Yu, 
%Chaos, Soliton \& Fractals 24 (2005) 1097-1104.
%\bibitem{43}
%I.~M.~Sokolov,  Phys.Rev. E 63 (2001) 056111.
%\bibitem{44}
%I.~S.~Gradshtein, I.~M.~Ryzhik, Tables of Integrals, Series, and
%Products, Acad. Press, New York, 1980.
%\bibitem{45}
%M.~Abramowitz, I.~A.~Stegun, Handbook of Mathematical Functions,
%Dover, New York, 1972.
%(2001) 389.
%\bibitem{46}
%G.~Doetsch, Introduction to the Theory and Application of the
%Laplace Transformations, Springer-Verlag, Berlin, 1974.
%\bibitem{48}
%K.~Diethelm, N.~J.~Ford, A.~D.~Freed, Nonlinear Dyn. 29 (2002) 3-22.
%\bibitem{49}
%P.~Arena, R.~Caponetto, L.~Fortuna, D.~Porto, 
%in Proceedings of ECCTD, Budapest, September 1997, 
%Technical University of Budapest, pp. 1259-1262.
%\bibitem{rz5}
%S.~G.~Samko, A.~A.~Kilbas, O.~I.~Marichev, ``Fractional Integrals and
%Derivatives and Some of Their Applications'', Nauka i Technika, Minsk
%(1987) (in Russian).

\end{thebibliography}
\end{document}